# Shaping the Future of Urban Mobility: Insights into Autonomous Vehicle Acceptance in Shanghai Through TAM and Perceived Risk Analysis

**Linxuan Yu[1, #], Miaomiao Shen[2, #], Jing Xu[3], Zihao Sang[4], Ruijia Li[5], Xiang Yuan[*]**

[1] School of Economics & Management, Shanghai Maritime University, Shanghai, China; yulinxuan@stu.shmtu.edu.cn

[2] School of Economics & Management, Shanghai Maritime University, Shanghai, China; 18757624686@163.com

[3] College of Mathematics and Physics, Shanghai University of Electric Power, Shanghai, 201306, China

xujing200210@163.com

[4] Merchant marine college, Shanghai Maritime University, Shanghai, China; 17367070173@163.com

[5] Ulster College at Shaanxi University of Science & Technology, Shanxi, China; 1535766542@qq.com

[*] Correspondence: School of Economics & Management, Shanghai Maritime University, Shanghai, China;

yuanx2109@163.com

[#] These authors contributed equally to this work.

# Abstract

Autonomous vehicles (AVs) have begun experimental commercialization initiatives in places such as Shanghai, China, and it is a valuable research question whether people's willingness to use AVs has changed from the prior. This study explores Shanghai residents' attitudes towards AVs by applying the Technology Acceptance Model (TAM), the Perceived Risk (BAR) model, and introducing perceived externalities as a new psychological variable. Through a survey in Shanghai, where AVs are operational, and structural equation modeling, it was found that perceived usefulness and ease of use positively influence willingness to use AVs, with perceived usefulness being the most significant factor. Perceived externalities have a positive impact, while perceived risk negatively affects willingness to use. Interestingly, ease of use increases perceived risk, but this is mitigated by the benefits perceived in usefulness. This research, differing significantly from previous studies, aims to guide government policy and industry strategies to enhance design, marketing, and popularization.

# 1. Introduction

If you want to imagine what the future of automobiles will look like, then autonomous vehicles (AVs) must be the best blueprint. If you want traffic to flow smoothly, free your hands, integrate smart technology, communicate with the vehicle anytime, anywhere, even if it is an online car without the distraction of bystanders, and realize interesting travel, AVs can do it all.

AVs are the hotspot of global development. In China, automobile enterprises represented by SAIC Group have been exploring the operation of L4-level AVs. In Beijing, Shanghai, Suzhou, Shenzhen and other areas, SAIC Robotaxi has begun to simulate commercialization scenarios for operational services, increasing horsepower to run out of driverless development acceleration. The public can get a fast car hailing experience through cell phone software, and the current driverless technology can already comfortably cope with the changing driving environment and guard passenger safety. If AVs are successfully applied to the market, they have the potential to dramatically change the transportation situation. AVs could reduce human-caused car accidents, make road travel more efficient, reduce carbon emissions, and contribute to environmental protection,and it also facilitates people who are incapable of driving automobiles for other reasons, such as physical fitness.(Fagnant & Kockelman,2015). However, the adoption of AVs also brings new risks, with the most common obstacles being security risks, privacy breaches and legal risks(Fagnant and Kockelman, 2015; Kyriakidis et al., 2015; Piao et al., 2016).

As the technology for AVs matures to the point where it is commercially viable, A wide range of stakeholders, including governments, automakers, and insurers, are intrigued by forecasting the future penetration of the market for AVs(Kim, Jae Hun, et al., 2022).Shanghai alone has China's first full-scenario driverless test site, as well as AVs in trial operation. However, consumer attitudes toward the emerging driverless cannot be known. Therefore, studying consumers' willingness to use AVs and the factors influencing it is crucial for stakeholders to determine a practical action plan for the future widespread use of driverless vehicles. This study will provide the government with a way to understand current consumer needs in terms of policy and other aspects, and automobile manufacturers will be able to understand consumer concerns in order to develop more appropriate production strategies and marketing tactics.

A number of recent studies have investigated this issue extensively. It was found that most users accept AVs a priori and want to use it ( Payre, et al.,2014 ; Bansal, et al.,2016 ; Krueger, et al.2016 ; Jardim, et al.,2013 ; Daziano, et al.,2017 ). At the same time these attitudes and underlying behavioral responses are likely to shift as the public becomes more aware of AVs and more and more technological lessons begin to make their way into the public sphere. For example, a majority of the population is initially reluctant to pay anything for advanced automation technology,however,as technology matures, and as they observe their neighborhoods and colleagues embrace AVs, they may change their perceptions of AVs (Bansal, et al.,2017 ). However Haboucha, et al. (2017) found that there are still some people who are very hesitant to adopt AVs, on the one hand due to cost considerations. On the other hand due to the perceived risks and fears of users (Peng Liu, et al., 2019; Bansal, et al., 2016; Kyriakidis, et al. 2015).

At the same time, there is individual heterogeneity in the willingness to use AVs, which has been found in several studies (Daziano, et al.,2017), as a result, a large number of experiments were carried out on the socio-demographic characteristics of the portion of the population choosing to own and use AVs in search of universal patterns. The first is in the study of regional and individual mass differences. Park, Jein, and Semi Han (2023) ; Hassan, Hany M. , et al. (2021) . investigated the preference of older adults to use AVs. Naderi, Hossein, and Habibollah Nassiri (2023) studied the performance situation of Iranians in terms of the degree of acceptance of AVs. Sener, et al. (2019) Selected the public of Texas to investigate the intention to use AVs. Bennett, et al. (2020) Investigated blind people's attitudes toward driverless vehicles (AVs) and their preference to travel in AVs. Studies have generally found that men, younger people, those with higher earnings, and qualifications are more inclined to use AVs ( Hohenberger, et al.,2016; Deb, et al.,2017; Hulse, et al.,2018 ).The demographic characteristics have been largely clarified. In addition, area of residence, size of residence, accident experience, and commuting mode all have a significant association with willingness to use Avs. Yoo Sunbin, and Shunsuke Managi (2021) It even tapped into the relationship between awareness of environmental protection and the willingness to purchase AVs.

Summarizing the above studies, it is not surprising to find that one of the common threads is that the intentions to use AVs not only varies from person to person, but also that the willingness to use them changes over time, geographically, and with the degree of development of AVs. It is a pity that due to the limitations of the development of AVs, very few AVs were running in the areas of previous studies. The subjects of these studies had limitations in their knowledge of AVs, and even more so, they did not really experience AVs running on the road.

At present, the development speed of AVs is incomparable, the user's willingness to use AVs may be undergoing a drastic change, and the research on AVs needs to develop a new version, so there are certain limitations in the previous research. Currently, Shanghai is already running L4 AVs on a trial basis, and the development of AVs is relatively mature. A new research study in Shanghai is necessary, and this study will involve people who actually have experience in AVs and investigate how the willingness of Shanghai residents to use them differs from the past.

Based on this, we conducted a survey based on the TAM and BAR models to investigate the willingness to use AVs in Shanghai, where AVs operate, and surveyed the people who have ridden in AVs. Then structural equation modeling was applied to explore the main factors affecting potential users' willingness to use AVs, so as to find the opportunity points for AVs development, aiming to suggest effective conclusions for AVs stakeholders.

The rest of the paper is organized as follows: section 2 presents the differences between this study and previous studies; section 3 presents the main hypotheses, model of this study, research methodology and data; then section 4 carries out the data analysis and modeling; section 5 details the results of the modeling analysis; and, finally, section 6 provides a discussion and conclusion.

## 2. Literature review

In the current body of research, researchers have been trying to find factors that perfectly summarize the factors that influence the willingness to use AVs ( Kim, et al.,2022 ). Hartwich, et al. (2019) found that the first ride experience was found to substantially increase users' trust and acceptance of AVs, and, after the first ride experience, trust and acceptance remained at a plateau. Kaye, et al. (2021) found that increases in trust and feelings of use were favorable to AVs' intentions to use, while age largely did not affect intentions to use by using meta-analysis.It is clear that these studies are limited to one specific aspect and do not rise to a systematic and complete system to fully explore the potential factors that lead to changes in intention to use AVs.

Through a review of previous research, the TAM model (Technology Acceptance Model) is a model that examines the factors that influence an individual's willingness to use a new technology, and it is one of the most popular models in use today, with perceived usefulness, perceived ease of use, and willingness to use as the main factors examined in the TAM model ( Davis, 1989). Some scholars will use TAM to predict consumer intentions towards AVs ( Rahman, et al.,2017) . Panagiotopoulos, et al. (2018) and Silvestri, et al.(2024) extended the original basic framework of TAM in the study found that perceived usefulness, perceived ease of use, perceived trust, and social influence were all useful predictors of behavioral intentions to own or use AVs, with perceived usefulness having the greatest impact. Buckley, er al. (2018) expanded on previous research on the level of trust in AVs, and the results show that subjective willingness and perceived behavior are key factors influencing trust.

Although these studies have linked influencing intentions to use AVs with perceived usefulness perceived ease of use, some of these TAM factors and their relationships remain controversial; for example, the technology acceptance model assumes that perceived ease of use affects perceived usefulness. However, the relationship between perceived ease of use and perceived usefulness has diverged in previous studies; Choi and Ji (2015) argued that perceived ease of use does not affect perceived usefulness. Lee, et al. (2019) argued that perceived ease of use increases when perceived usefulness also increases. Generally, the user finds the use of AVs difficult to master, and he may not feel the usefulness and utility of this technology, which reduces the frequency of their use of AVs. Therefore, our study hypothesizes that perceived ease of use has a positive relationship on perceived usefulness.

In addition, the relationship between perceived risk and other influences is not yet known. Lee, et al. (2019) study details how perceived ease of use and perceived usefulness influence intention to use and finds that perceived risk influences intention to use, but he finds that other factors (perceived usefulness and perceived ease of use) have no influence on perceived risk. This implies that users' perceived risk of AVs is only related to non-human factors, such as vehicle breakdowns or natural disasters. And reviewing previous studies, we found that users tend to prefer configurations such as steering wheel or brakes that can be retained in AVs, which is to increase the usefulness to reduce the likelihood of accidents. On the contrary, if the AVs are empty, losing the ease of use, and the driving is completely under the control of the vehicle, users will feel unsafe. Therefore, we propose the more reasonable hypothesis that perceived ease of use has a significant positive effect on perceived risk, while perceived usefulness negatively affects perceived risk.

Nowadays, there are also a considerable number of scholars who will include perceived risk among the influencing factors in their research ( Liu2, et al., 2019 ; Lee et al., 2019) . Therefore, this study also refers to the BRA model (benefit-risk analysis model), which is one of the mainstream frameworks for consumer behavior research, in building the model, which considers consumers' perceived risk and perceived benefit as determinants of willingness ( Yingyu Zhang, 2015 ). In addition, it has also been argued that the perception of environmental effects generated by Avs also affects willingness to use and attitudes ( Shabanpour, Ramin et al. 2018 ). According to the above-mentioned theories, this study introduces perceived externalities as the main variables and analyzes the factors influencing the willingness to use AVs based on perceived usefulness, perceived ease of use in the TAM model and perceived risk in the BRA model.

Another limitation of the above studies is that the above literature studies mainly focused on potential users without actual road ride experience, and the sample group in the survey had no experience of driving or riding AVs (Liu Liu1, et al., 2019 ), with some using driver simulators ( Buckley, et al., 2018; Körber, et al., 2018) , some are in test road driving ( Liu2, et al., 2019 ). While simulators and experimental roads are safer, at the same time the drawbacks are more pronounced. Heterogeneity may also exist due to preferences or concerns between people. More specifically, a single factor may have different effects on preferences or concerns about AVs. In this study, we conducted a survey in Shanghai on people who have experienced driverless car rides, which provided us with a unique survey data resource as Shanghai has the richest comprehensive test demonstration area for intelligent networked vehicles in China in terms of test sites, and there are automobile companies, such as SAIC and Azure, that have conducted pilot operations of AVs in Shanghai.

In summary, at present, scholars at both domestic and international level have mainly analyzed the factors that can be regarded as influencing the intention of driverless AVs, but they have mainly focused on potential passengers without actual riding experience, and there is a lack of research in the face of driverless AVs that have been operated on the road. A few of these scholars have utilized the TAM model to analyze the associations between the influencing factors. However, due to the limitations of the original model, the influencing factors considered by TAM do not fully cover the characteristics of AVs, and there are still some issues that have not been properly addressed in the existing studies regarding the relationship between the influencing factors of intention to use.

Therefore, in the face of these gaps, there are three innovations in this study. One lies in the sample data, the sample data of this study comes from the Shanghai city with driverless AVs operation, previous studies have not done a comprehensive survey and analysis for the areas with L4 driverless operation, so for the sake of the breadth and reliability of the research results, it is necessary to establish a survey that covers the group of people who have experience of AVs. The second lies in the modeling method, this study extends and supplements the original TAM model, introduces the risk perception (BAR) model, and adds the element of perceived externalities. This study developed a questionnaire and constructed a structural equation model based on four influencing factors: perceived risk, perceived usefulness, perceived ease of use, and perceived externality. Third, we found path relationships that differed from previous studies. In previous studies, perceived risk was considered to have an isolated effect, while our study found that perceived ease of use positively affect on perceived risk, while perceived usefulness negatively affects perceived risk.

# 3. Theoretical framework and hypotheses

## 3.1 Perceived Ease of Use and Perceived Usefulness

The foundational Theory of Acceptance Model (TAM) posited that perceived ease of use positively influences perceived usefulness, and both significantly impact the usage intention (Davis, 1989; Davis, et al., 1989). Most AV-related studies employing the TAM have confirmed these basic assumptions (Lee et al., 2019). Typically, when AVs are perceived as easy to use, their perceived usefulness is enhanced, thereby strengthening the intention to use them (Xu et al., 2018; Lee et al., 2019; Panagiotopoulos, et al., 2018). Hence, the following hypotheses are proposed:

**H1.** Perceived ease of use significantly positively affects perceived usefulness.

**H2.** Perceived ease of use significantly positively affects the intention to use.

**H3.** Perceived usefulness significantly positively affects the intention to use.

## 3.2 Perceived Externalities

Perceived externalities refer to the impact of a decision or action on those other than the decision-maker or actor (Helbling and Thomas, 2010). In the context of AVs, environmental and traffic-related externalities can influence the intention to use (Shabanpour, Ramin, et al., 2018; Yiqing Tang, 2020).

Previous research indicates that users' perceptions of a product's cognitive and utility factors influence their assessment of the product's positive externalities, impacting their intention to use and purchase behavior (Bartels, et al., 2011).

Enhanced perceived usefulness and ease of use in AV technology can make users believe that AVs are easy to use and match their interaction preferences, reducing operational errors and the risk of driver fatigue, thereby decreasing traffic accidents. Furthermore, promoting the adoption of AV technology can lead to more positive externalities. Therefore, the following hypotheses are proposed:

**H4.** Perceived ease of use significantly positively affects perceived externalities.

**H5.** Perceived usefulness significantly positively affects perceived externalities.

**H6.** Perceived externalities significantly positively affect the intention to use.

## 3.3 Perceived Risk

The BRA model posits that perceived risk influences the intention to use. According to the theory of perceived risk, consumers engage in risk-bearing behaviors in consumption activities, often based on the principle of minimal perceived risk ([Bauer, Raymond A, 1960](#)). Furthermore, [Huang (2013)](#) found that the willingness to use AVs is influenced by motivations such as novelty, driving experience, and enjoyment. However, the mechanism by which driving pleasure and experience influence the intention to use is not conclusively determined. Some scholars argue that higher levels of vehicle autonomy, i.e., greater perceived ease of use, reduce driving pleasure, increasing the intention to use ([Huang, Tianyang, 2021](#)). Others suggest that despite interest in AVs, consumers may be more cautious in their attitudes towards them ([Deichmann, Johannes, 2023](#)). However, there is a general belief that vehicles should provide a sense of driving and control ([Krishna, G, 2021](#)). Therefore, this study hypothesizes that as perceived ease of use increases, driving pleasure may decrease, and the lack of control devices (like steering wheels), though convenient, might also lower the sense of control over the vehicle, thereby increasing perceived risk.

Previous research suggests that perceived benefits influence public perception, and if the perceived usefulness of AVs is high and considered to have positive externalities, this may lower perceived risk ([Chamata, 2018](#)). Therefore, this study posits that higher perceived usefulness and externalities might reduce the perception of potential risks, leading to the following hypotheses:

**H7.** Perceived ease of use significantly positively affects perceived risk.

**H8.** Perceived usefulness significantly negatively affects perceived risk.

**H9.** Perceived externalities significantly negatively affect perceived risk.

**H10.** Perceived risk significantly negatively affects the intention to use.

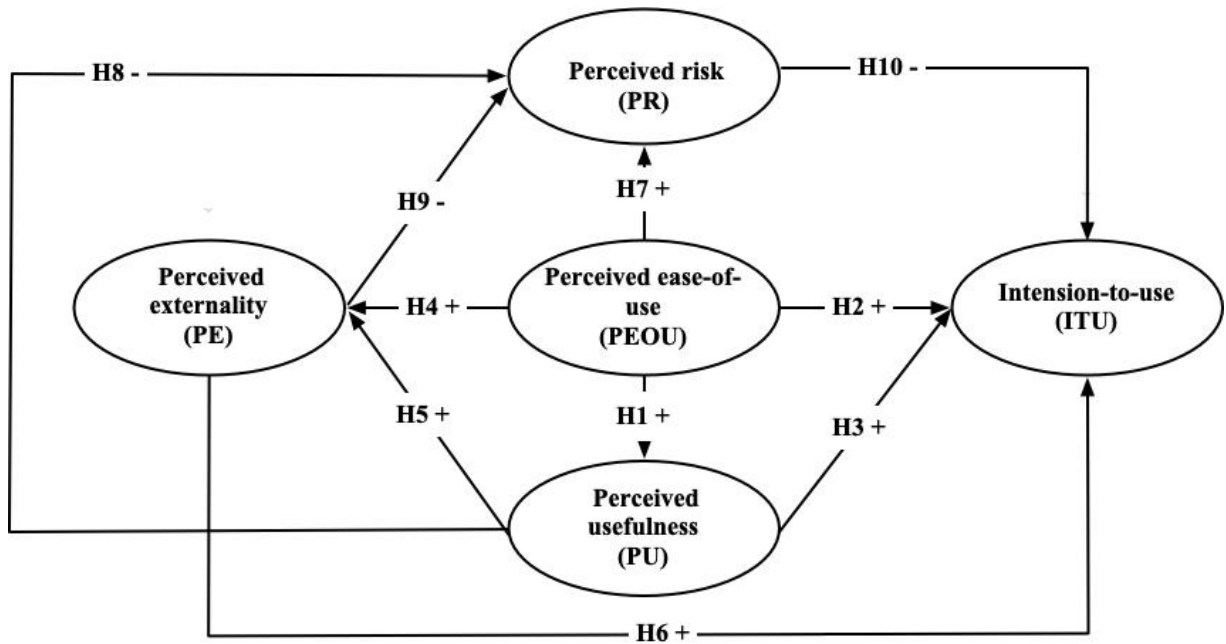

note: ***: p<0.01, **: p<0.05, *: p<0.1

Fig.1 illustrates the interrelationships between all the research hypotheses and the individual factors.

**Fig.1 Research model.**

## 3.4 Research Methodology and Data Collection

This survey was conducted from January to March 2023, focusing on residents of various districts in Shanghai, China. The data collection was carried out through an online questionnaire survey via the renowned Chinese survey platform,

"Wenjuan.com (www.wenjuan.com)." To ensure the objectivity and scientific validity of the survey, and to enhance its universality and representativeness, the following methods were adopted:

(1) Stratified sampling was used to divide Shanghai into different districts. The theoretical distribution of questionnaires was calculated proportionally based on the permanent population figures of each district from the Shanghai Statistical Yearbook 2021.

(2) Sample communities were further extracted within each district. The sample size for each community was determined based on the number of communities in each district and the proportionate allocation of theoretical sample sizes, using weightings calculated from these figures. The selection of sample communities was randomly conducted using a Python program.

(3) An equal number of questionnaires were distributed to each selected community through an online survey. In total, 1000 questionnaires were distributed and collected, with 144 invalid responses (including blank and inconsistent responses) excluded, resulting in an effective response rate of 85.6%.

To test the research hypotheses, the questionnaire was designed with specific measurement indicators based on metrics found in previous literature. The latent variables were measured using a Likert five-point scale, as shown in Table 1. Additionally, since the AVs studied in this article are at Level 4, which is a high degree of automation (SAE, 2014), respondents were required to watch a video defining Level 4 AVs before answering the questionnaire, ensuring the authenticity and effectiveness of their responses. Demographic characteristics of the valid respondents are presented in Table 2.

**Table1. Constructs and questionnaire items.**

| Constructs | Items | Sources |
|---|---|---|
| Perceived risk (PR) | I am concerned that the computer systems of AVs are hacked (PR1) | Liu1 et al. (2019); Featherman and Pavlou (2003); Liu3 et al. (2019); Lee et al. (2019) |
| | I am concerned about the legal liability of drivers or owners of | |
| | AVs (PR2) | |
| | I am concerned about AVs travelers' privacy disclosure (PR3) | |
| **Perceived externality (PE)** | AVs can reduce traffic crashes (PE1) | Liu1 et al. (2019); Anderson et al. (2014); Bansal et al. (2016); Fagnant and Kockelman (2015); Howard and Dai (2014); Liu3 et al. (2019) |
| | AVs can reduce vehicle emissions and pollution (PE2) | |
| | AVs can reduce traffic congestion (PE3) | |
| **Perceived ease-of-use (PEOU)** | Learning to operate an AVs would be easy for me (PEOU1) | Davis (1989); Davis and Venkatesh (1996); Lee et al. (2019) |
| | Interacting with an AVs would not require much mental effort (PEOU2) | |
| | It's easy to get a AVs to do what I tell it to do (PEOU3) | |
| | I would find an AVs is easy to use (PEOU4) | |
| **Perceived usefulness (PU)** | Using an AVs is very useful in life and work (PU1) | Davis (1989); Davis and Venkatesh (1996); Lee et al. (2019) |
| | Using an AVs can make commuting and travelling more efficient (PU2) | |
| | Using an AVs can improve travelling comfort (PU3) | |
| **Intension-to-use (ITU)** | Given access to an AVs, my intention to utilize it would be present (ITU1) | Davis (1989); Davis and Venkatesh (1996); Lee et al. (2019) |
| | I anticipate using an autonomous vehicle system (AVS) in the future (ITU2). | |

Table 2. Demographic information of participants in Shanghai City (n=856)

| Measure | Value | Frequency (Percentage) |
|---|---|---|
| **Gender** | Male | 499 (58.34) |
| | Female | 357 (41.66) |
| **Age** | 18 and below | 11 (1.29) |
| | 18-22 | 161 (18.78) |
| | 22-30 | 423 (49.41) |
| | 30-45 | 236 (27.58) |
| | 45 and above | 25 (2.93) |
| | High school and below | 37 (4.34) |
| Education | Junior college education | 159 (18.54) |
| | undergraduate education | 600 (70.07) |
| | Postgraduate and above | 60 (7.04) |
| | 105000 and below | 367 (42.84) |
| Annual income | 105000-700000 | 417 (48.71) |
| | 700000 and above | 72 (8.45) |
| | Professionals (doctors, lawyers, etc.) | 157 (18.31) |
| | clerical staff | 146 (17.02) |
| | Freelance, private enterprise, self-employed | 249 (29.11) |
| | government employee | 79 (9.27) |
| Occupation | student | 120 (13.97) |
| | mechanic | 46 (5.40) |
| | Unemployed/retired | 20 (2.35) |
| | Teachers | 34 (3.99) |
| | Other | 5 (0.59) |

# 4. Structural Equation Model Analysis and Validation

## 4.1 Structural Equation Model

To quantitatively describe the causal relationships among the five latent variables set in this study and explore the impact of Perceived Risk (PR), Perceived Enjoyment (PE), Perceived Ease-of-Use (PEOU), and Perceived Usefulness (PU) on the Intention-to-Use (ITU) of AVs, this paper employs Structural Equation Modeling (SEM). SEM is a statistical analysis method used to explore complex relationships among variables (Ullman et al., 2012). It estimates the relationships between multiple latent variables and their observed variables simultaneously. These relationships are represented as structural equations, with arrows indicating causal links between variables.

SEM consists of two components: the measurement model and the structural model. The measurement model estimates latent variables using observed variables and parameters representing relationships between latent variables. The structural model estimates causal relationships between variables, including parameters indicating the relationships and error terms. The specific expressions are as follows:

$$\eta = B\eta + \Gamma\xi + \zeta \#(1)$$

$$Y = \Lambda_y + \varepsilon \#(2)$$

$$X = \Lambda_x + \sigma \#(3)$$

$\xi$ represents exogenous latent variables, $\eta$ represents endogenous latent variables, $X$ and $Y$ are their observed variables; $\Lambda_x$ and $\Lambda_y$ are factor loadings between $X$ and $\xi$, and $Y$ and $\eta$, respectively; $\sigma$ and $\varepsilon$ represent measurement errors for $X$ and $Y$; $\zeta$ is the disturbance term unexplained by the model; $B$ is the matrix of coefficients between endogenous latent variables, and $\Gamma$ represents the matrix of coefficients between exogenous latent variables.

## 4.2. Model Testing

Before testing path relationships using SEM, it is common to validate the sample data and the model to ensure the model's fit.

### 4.2.1. Reliability Testing

For the questionnaire, reliability was assessed using the Cronbach's α coefficient. The specific reliability test results are shown in Table 3. The Cronbach's α for ITU is 0.663, indicating good reliability. PR (0.766), PE (0.765), PU (0.776), and PEOU (0.823) all have Cronbach's α above 0.7, suggesting good reliability for these dimensions as well. Overall, the scale exhibits excellent reliability, with a Cronbach's α of 0.919. Referring to the suggested values of Cronbach's α in previous literature ([Hair et al., 2014;](#) [Fornell et al., 1981](#)), the measurement indicators of the questionnaire are reliably established.

**Table 3 - Reliability Tests**

| Constructs | N   | Cronbach'sα |       |
|------------|-----|-------------|-------|
| **PR**     | 826 | 0.766       |       |
| **PE**     | 826 | 0.765       |       |
| **PEOU**   | 826 | 0.823       | 0.919 |
| **PU**     | 826 | 0.776       |       |
| **ITU**    | 826 | 0.663       |       |

## 4.2.2. Construct Validity

Confirmatory factor analysis was conducted on both the measurement and structural models to analyze model fit. The critical values and actual values of the fit indices are shown in Table 4. The model's chi-square to degrees of freedom ratio is less than 3, and both GFI and AGFI, which indicate the extent to which the theoretical model explains covariance and variance, are above 0.90 ([Anderson and Gerbing, 1988](#)). Comparative Fit Index (CFI) results are above 0.90, Root Mean Square Error of Approximation (RMSEA) is below 0.10, Tucker-Lewis Index (TLI) is above 0.90, and Root Mean Square Residual (RMR) is below 0.05. Other fit indices are within critical ranges, indicating good model fit for path analysis ([Bentler and Bonett, 1980;](#) [Anderson and Gerbing, 1988;](#) [Hatcher et al., 2013;](#) [Dion, 2008;](#) [Lee et al., 2008;](#) [Lee et al., 2009;](#) [Kim et al., 2011](#)).

**Table 4 - Model Fit Tests**

| $\chi^2/df$ | GFI | RMSEA | RMR | CFI | NFI | NNFI | AGFI | TLI |
|---|---|---|---|---|---|---|---|---|
| <3 | >0.9 | <0.10 | <0.05 | >0.9 | >0.9 | >0.9 | >0.9 | >0.9 |
| **1.667** | 0.979 | 0.029 | 0.018 | 0.990 | 0.976 | 0.960 | 0.968 | 0.987 |

## 4.2.3. Convergent Validity

Confirmatory factor analysis, as shown in Table 5, reveals that the Squared Multiple Correlations (SMC) for each variable are greater than 0.36. For Composite Reliability (CR), except for ITU which is slightly lower (0.663) but still acceptable, all other variables have a CR above 0.7 (Hair et al., 2014). Additionally, for model validity, each observed variable's standardized loading (S. Loading) is above 0.5 (Hair et al., 2014). As for Average Variance Extracted (AVE), except for ITU which is slightly below 0.5 (0.497), all other variables are above 0.5 (Hair et al., 2014). These indicate that the model's convergent validity meets the requirements (Fornell and Larcker, 1981; Chin, 1998) and the results of the model are reliable (Gefen et al., 2000; Hair et al., 2017).

**Table 5. validity assessment**

| Constructs | Measure | S. Loading | S.E. | t | P | SMC | CR | AVE |
|---|---|---|---|---|---|---|---|---|
| | PR1 | 0.759 | - | - | - | 0.576 | | |
| **PR** | PR2 | 0.669 | 0.051 | 17.708 | *** | 0.448 | 0.769 | 0.527 |
| | PR3 | 0.746 | 0.054 | 19.999 | *** | 0.557 | | |
| **PE** | PE1 | 0.737 | - | - | - | 0.543 | 0.766 | 0.522 |
| | PE2 | 0.696 | 0.048 | 19.448 | *** | 0.484 | | |

|  | | | | | | | |
|---|---|---|---|---|---|---|---|
|  | PE3 | 0.733 | 0.049 | 20.54 | *** | 0.537 |  |  |
|  | PU1 | 0.707 | - | - | - | 0.5 |  |  |
| **PU** | PU2 | 0.727 | 0.055 | 19.219 | *** | 0.529 | 0.776 | 0.536 |
|  | PU3 | 0.762 | 0.056 | 20.246 | *** | 0.581 |  |  |
|  | PEOU1 | 0.743 | - |  | - | 0.552 |  |  |
| **PEOU** | PEOU2 | 0.701 | 0.049 | 19.691 | *** | 0.491 | 0.823 | 0.538 |
|  | PEOU3 | 0.74 | 0.048 | 20.833 | *** | 0.548 |  |  |
|  | PEOU4 | 0.748 | 0.048 | 21.194 | *** | 0.56 |  |  |
| **ITU** | ITU1 | 0.69 | - |  | - | 0.476 | 0.663 | 0.497 |
|  | ITU2 | 0.719 | 0.088 | 11.649 | *** | 0.517 |  |  |

note: ***: p<0.01, **: p<0.05, *: p<0.1.

## 4. 3. Path Relationship Testing

Having passed tests for structural and convergent validity, the model demonstrates good fit and is ready for path analysis. Path analysis allows for the testing of hypothesis significance. Results, as shown in Table 6 and Figure 1, indicate that H1, H2, H6, and H10 are significant at the 1% level, while H3, H7, and H8 are significant at the 5% level. Other hypotheses are rejected at the 5% level. Specifically, standardized path coefficients (S. Coefficient) reflect the magnitude of influence among variables. Perceived Ease-of-Use (0.396), Perceived Usefulness (1.677), and Perceived Externalities (0.584) all significantly and positively influence the Intention-to-Use AVs, validating hypotheses H2, H3, and H6. Perceived Risk, with a coefficient of -0.46, significantly negatively affects the Intention-to-Use, thereby validating hypothesis H10. Among the factors, Perceived Usefulness exerts the most substantial impact on Intention-to-Use, while Perceived Ease-of-Use has the minimal influence. Additionally, Perceived Ease-of-Use has a positive influence on Perceived Usefulness (0.92), validating H1. Perceived Ease-of-Use positively influences Perceived Risk (0.599), while Perceived Usefulness negatively influences it (-1.064), confirming H7 and H8.

**Table 6 - PATH Coefficient Estimates**

| Hypothesis | Path relation | | | S. Coefficient | P |
|---|---|---|---|---|---|
| **H1** | **PEOU** | → | **PU** | **0.92** | **0.000*** |
| **H2** | **PEOU** | → | **ITU** | **0.396** | **0.003*** |
| **H3** | **PU** | → | **ITU** | **1.677** | **0.012** |
| H4 | PEOU | → | PE | 0.324 | 0.316 |
| H5 | PU | → | PE | -0.889 | 0.454 |
| **H6** | **PE** | → | **ITU** | **0.584** | **0.000*** |
| **H7** | **PEOU** | → | **PR** | **0.599** | **0.020** |
| **H8** | **PU** | → | **PR** | **-1.064** | **0.016** |
| H9 | PE | → | PR | 1.315 | 0.546 |
| **H10** | **PR** | → | **ITU** | **-0.46** | **0.009*** |

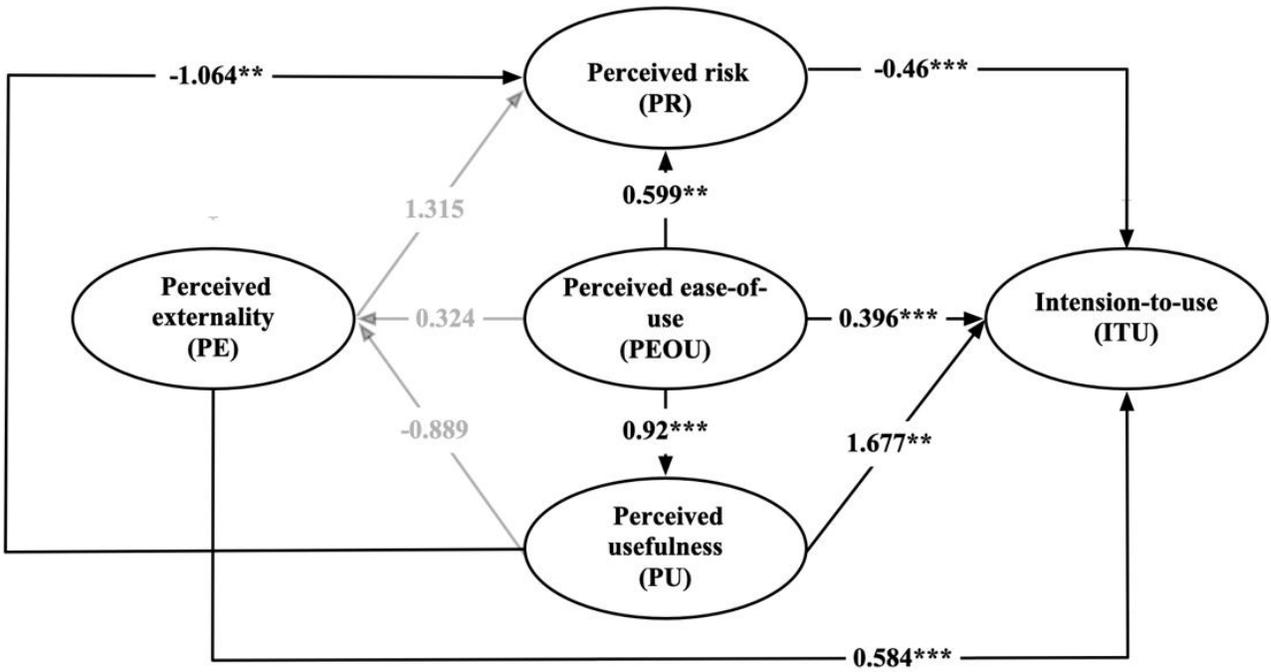

**Fig.2 Estimated structural equation model.**

# 5. Results Analysis

(1) For AVs, Usefulness is More Important than Ease-of-Use

Past studies have shown varying opinions on whether Perceived Usefulness and Perceived Ease-of-Use both affect the Intention-to-Use. Beyond the original Technology Acceptance Model (TAM) (Davis, 1989; Davis et al., 1989), reports by Panagiotopoulos and Dimitrakopoulos(2018) accepted these assumptions, while Hein et al. (2018) found no significant impact of Perceived Ease-of-Use on the intention to use. Despite varying conclusions in past research, a prevailing view is that Perceived Usefulness has the greatest impact on the Intention-to-Use AVs, relatively more than Perceived Ease-of-Use ([Hein et al. 2018](); [Panagiotopoulos 和 Dimitrakopoulos, 2018](); [Lee et al., 2019]()). According to the results in Table 6, the order of impact on Intention-to-Use from most to least significant is: Perceived Usefulness, Perceived Externalities, Perceived Risk, and Perceived Ease-of-Use (all significantly affecting Intention-to-Use). It is evident that users are more likely to accept and use the technology when they perceive tangible benefits of AVs in their lives or work.

Additionally, the research discovered that Perceived Ease-of-Use has a positive effect on Perceived Usefulness, suggesting that improvements in the Perceived Ease-of-Use of autonomous vehicle technology could enhance its perceived utility. This suggests that when users find AV technology easy, convenient, and user-friendly, they are more likely to appreciate its practicality and utility, thus strengthening their intention to use it.

(2) Intention-to-Use AVs Influenced by Psychological Factors

The Intention-to-Use AVs is influenced by Perceived Externalities [(Choi and Ji, 2015](); [Hohenberger et al.]() ,2017;[Liu3 et al.,2019]()) and Perceived Risk ([Ward et al.,2017](); [Xu et al.,2018](); [Liu3 et al., 2019]()), as supported by previous studies. Specifically, when users believe that using AVs will benefit the environment and society, their Intention-to-Use increases. Similarly, when AVs are perceived as safe, the intention to use them strengthens. In this study, significant standard path coefficients for Perceived Risk and Perceived Externalities substantiate this view.

Past research has proposed strategies to effectively mitigate Perceived Risk and enhance Perceived Externalities of AVs. For instance, increasing the anthropomorphic features of AVs and designing personalized avatars could enhance users' trust in AVs ([Waytz et al., 2014]()). [Liu1 et al. (2019)]() suggested promoting the social utility and effects of AVs on social media to heighten users' perception of AVs' social externalities.

(3) Greater Perceived Usefulness and Ease-of-Use of AVs Can Enhance Perceived Safety

Contrary to some past studies, [Lee et al., (2019)]() posited that Perceived Risk has an isolated effect, with Perceived Ease-of-Use and Perceived Usefulness not significantly impacting Perceived Risk. This paper finds that Perceived Ease-of-Use has a significant positive impact on Perceived Risk, while Perceived Usefulness negatively influences it.

The increase in Perceived Ease-of-Use leading to heightened Perceived Risk might be due to firstly, an overly simple user experience making consumers believe the technology is prone to loss of control, thus increasing Perceived Risk. Secondly, a simplified operation experience often deprives users of the sense of control over the vehicle, for instance, the absence of a steering wheel or other control devices, which can elevate users' Perceived Risk. Therefore, it's advisable to retain some control devices and find a balance in operational complexity during the design process. Moreover, the significant negative relationship between Perceived Usefulness and Perceived Risk suggests that enhancing users' Perceived Usefulness can reduce their Perceived Risk. This is often because when users perceive AV technology as more convenient and beneficial, they might underestimate the potential risks involved, thereby reducing their perceived risk levels. Thus, in promoting and applying AV technology, it's important to focus on increasing Perceived Usefulness, enabling users to better recognize the benefits and advantages of the technology, consequently reducing their Perceived Risk.

## 6. Discussion and Conclusion

With the gradual commercialization of AVs, this study conducted field research in Shanghai, where AVs operate. After collecting responses to a questionnaire from 1,000 Shanghai residents, the intrinsic relationship between the willingness to use AVs and the factors influencing them was investigated by combining the TAM and BAR models and analyzing the data using structural equation modeling. The analyzed results satisfy the requirements of reliability, validity, and convergence.User's needs and concerns of using emerging AVs can be perceived by policy makers and car manufacturers through this study, and the government and car companies can better understand the public's concerns and the influencing factors of the public's willingness to use, to explore measures that can better promote the full and deep implementation of driverless.

Since the samples of previous studies have never covered Shanghai and AVs have begun to develop trends in Shanghai, users' willingness to use them may have shifted. This study investigated the attitudes of Shanghai residents' perceptions of AVs from multiple dimensions.Our survey conveys that more than half of the respondents' attitudes towards AVs are above neutral. This indicates that overall, public attitudes are positive, but not as positive as in previous studies by scholars in China ( Schoettle, et al.,2014 ). Nonetheless, consistent with previous research, the attitudes of Chinese residents toward the perception of driverless technology have been more optimistic than in other regions ( Li, et al.,2023 ). In addition, the results indicated that the public has an initial market awareness of AVs, but there is still much room for improvement in the public's direct perception of AVs due to the fact that they have not been commercialized on a large scale. This is similar to the findings of previous studies that Kyriakidis, et al. (2015) found 3.69% of respondents out of 5,000 questionnaires stated that driverless will take less than 50 years to reach 50% market share; Bansal, et al. (2016) predicted that driverless will be adopted by Americans in the long term. Previous studies have also focused heavily on factors of concern about AVs, such as privacy breach issues, legal issues, and safety concerns ( Kyriakidis, et al.,2015 ; Liljamo, et al.,2018 ), the findings of this study similarly found that the main concern of the worried respondents and some of the neutral respondents was the safety of AVs. This finding was echoed in Zoellick, et al. (2019 ) research can also be corroborated.

  Therefore, in order to increase the awareness of AVs and reduce users' doubts and concerns about AVs, combined with the survey data, we believe that companies can increase their promotional efforts, which is supported by existing research, the Talebia, et al. (2018) of research demonstrating that marketing can accelerate the diffusion of AVs. However it is not enough to focus on the impact of AVs on people, there also needs to be a focus on the importance and concerns about the safety of the technology ( Hudson, et al.,2019 ). Talebia, et al. (2018) also stated that the effect of marketing is limited and individual satisfaction plays a key role in deciding whether to use AVs or not. Therefore, car companies should continue to be concerned about the technical safety of AVs, such as focusing on the system stability of AVs during driving, the accuracy of external environment perception, and the security of data transmission and use to comprehensively improve the user's confidence in the use of AVs. In addition, car companies should accelerate data collection and testing under complex road conditions, strengthen safety verification, and improve technical safety in response to problems that arise during testing. Before the commercialization of driverless technology, the government also needs to be closely involved. On the one hand, from the perspective of AVs themselves, the Mordue et al. (2020) study found that government regulation could have a huge impact on the pace of automotive innovation and the establishment of a system that is consistent with government values; from the public's perspective, government regulation could also help to alleviate some of the public's concerns about the potential dangers of AVs, such as loss of control, and therefore legislation is needed to address this issue. On the other hand, enterprises have to realize breakthrough innovations on safety issues, which invariably requires the government to implement the whole industry chain investment and policy subsidies for AVs.

This study complements and extends the original model and finds different conclusions from previous studies. By combining two models, TAM and BRA, we set up four dimensions, perceived risk, perceived externality, perceived ease of use, and perceived usefulness, and used SEM to explore whether they correlate with consumers' desire to use them. The results of the paths are all consistent with our hypotheses. First, we find that perceived ease of use is significantly positive for perceived usefulness. This result is consistent with Choi and Ji (2015) 's findings, contrary to the findings of Lee, et al. (2019) whose findings are similar. However Lee, et al. (2019) study found that perceived risk has an effect of insularity and that perceived usefulness and perceived ease of use are not remarkably associated with perceived risk. In this study, we found that perceived ease of use is remarkably positive on perceived risk and on opposite, perceived usefulness is remarkably negative on perceived risk. This suggests that if the maneuvering of a driverless car is too simple instead, it will increase the consumer's perceived risk of the vehicle; enhancing the value of the vehicle's usefulness in the consumer's mind can reduce the consumer's perceived risk. This reaffirms the validity of the automaker's approach of retaining some manual controls, as potential users' concerns about the safety of AVs lead them to choose to retain the steering wheel of a conventional car as a means of reducing their risk perception.

Compared with these previous studies, this study additionally considered the influence of environmental effects on the perceived attitudes of AVs by introducing the variable of perceived externality, which multidimensionally examines the complete chain relationship of willingness to use. The results novelty found that perceived ease of use and perceived usefulness are significantly positive to perceived externalities, which suggests that driverlessness can reduce the risk of driver error and fatigue driving when it makes users feel that it is practically helpful or beneficial in their life or work, thus reducing the occurrence of traffic accidents and promoting more positive externalities. In terms of externalities that affect potential users' willingness to use them, AVs can reduce traffic accidents, reduce environmental pollution, and reduce traffic congestion. Governments can focus on maximizing social benefits from AVs and avoiding negative externalities by recommending comprehensive and long-term supportive investment in their impacts (Mordue et al.,2020). Nazari, et al. (2018) also stated in their studies that the government can promote green travel modes and adopt proactive policies before the commercialization of AVs to eliminate public concerns and enhance externalities. And car companies can gain public trust and concern about AVs by focusing on the social value of AVs, such as promoting the aspects that AVs can save energy, reduce carbon emissions, and reduce human traffic accidents.

Nordhoff et al. (2018) ; Paddeu et al. (2020) investigated users' acceptance to automated shuttle after riding and the result is they are more likely to use AVs when their previous experience with them has been favorable. Xu et al. (2018) also found that the ride experience can increase trust and perceived usefulness as well as perceived ease of use. So companies can expand the scope of AVs test drives to make potential users experience AVs and further increase willingness to use and trust after the experience, forming a virtuous cycle.

Of course, this study has limitations. Firstly, since only L4 AVs are currently in trial operation in Shanghai, our survey focuses only on L4 AVs. Second, our survey targets Shanghai residents, which may create the problem of insufficient representativeness and generalizability. As cultural practices vary from district to district, they may hold very different views. In addition, different countries vary in their level of development of AVs, and residents living in different countries may present quite different differences in their perceptions of AVs. Therefore, if the questionnaire is conducted in different countries, it may produce different results. Finally, we have only explored four influencing factors: perceived risk, perceived externalities, perceived ease of use, and perceived usefulness; however, willingness to use may be affected by many underlying restraints.

The three points will be taken into account in future studies. First, future research could separately investigate user acceptance of different levels of AVs, including the L1-L5, where different levels of automation may lead to differences in users' willingness to use them. Second, we can try to conduct questionnaire surveys in multiple countries to collect more diverse data to complement the existing data to improve our model. Finally, future research can consider more factors that can influence the willingness to use AVs, and can also explore the changes of these factors over the time span.

# Funding

This research received no external funding.

# Data Availability Statement:

Data are available from authors upon reasonable request.

# Conflicts of Interest:

The authors declare no conflicts of interest.

# References

[1] On-Road Automated Driving (ORAD) Committee. Taxonomy and definitions for terms related to on-road motor vehicle automated driving systems. SAE International, 2014.

[2] Liu1, Peng, et al. "Willingness to pay for self-driving vehicles: Influences of demographic and psychological factors." Transportation Research Part C: Emerging Technologies 100 (2019): 306-317.

[3] Fagnant, Daniel J., and Kara Kockelman. "Preparing a nation for autonomous vehicles: opportunities, barriers and policy recommendations." Transportation Research Part A: Policy and Practice 77 (2015): 167-181.

[4] Piao, Jinan, et al. "Public views towards implementation of automated vehicles in urban areas." Transportation research procedia 14 (2016): 2168-2177.

[5] Kyriakidis, Miltos, Riender Happee, and Joost CF de Winter. "Public opinion on automated driving: Results of an international questionnaire among 5000 respondents." Transportation research part F: traffic psychology and behaviour 32 (2015): 127-140.

[6] Kim, Jae Hun, et al. "Determinants of personal concern about autonomous vehicles." Cities 120 (2022): 103462.

[7] Park, Jein, and Semi Han. "Investigating older consumers' acceptance factors of autonomous vehicles." Journal of Retailing and Consumer Services 72 (2023): 103241.

[8] Hassan, Hany M., et al. "Older adults and their willingness to use semi and fully autonomous vehicles: A structural equation analysis." Journal of transport geography 95 (2021): 103133.

[9] Naderi, Hossein, and Habibollah Nassiri. "How will Iranian behave in accepting autonomous vehicles? Studying moderating effect on autonomous vehicle acceptance model (AVAM)." IATSS Research 47.4 (2023): 433-446.

[10] Sener, Ipek N., Johanna Zmud, and Thomas Williams. "Measures of baseline intent to use automated vehicles: A case study of Texas cities." Transportation research part F: traffic psychology and behaviour 62 (2019): 66-77.

[11] Bennett, et al. "Willingness of people who are blind to accept autonomous vehicles: An empirical investigation." Transportation research part F: traffic psychology and behaviour 69 (2020): 13-27.

[12] Payre, et al. "Intention to use a fully automated car: Attitudes and a priori acceptability." Transportation research part F: traffic psychology and behaviour 27 (2014): 252-263.

[13] Bansal, Prateek, Kara M. Kockelman, and Amit Singh. "Assessing public opinions of and interest in new vehicle technologies: An Austin perspective." Transportation Research Part C: Emerging Technologies 67 (2016): 1-14.

[14] Krueger, Rico, Taha H. Rashidi, and John M. Rose. "Preferences for shared autonomous vehicles." Transportation research part C: emerging technologies 69 (2016): 343-355.

[15] Haboucha, Chana J., Robert Ishaq, and Yoram Shiftan. "User preferences regarding autonomous vehicles." Transportation Research Part C: Emerging Technologies 78 (2017): 37-49.

[16] Jardim, Adam Sebastian, Alex Michael Quartulli, and Sean Vincent Casley. "A study of public acceptance of autonomous cars." Worcester Polytechnic Institute: Worcester, MA, USA (2013): 156.

[17] Daziano, Ricardo A., Mauricio Sarrias, and Benjamin Leard. "Are consumers willing to pay to let cars drive for them? Analyzing response to autonomous vehicles." Transportation Research Part C: Emerging Technologies 78 (2017): 150-164.

[18] Liljamo, Timo, Heikki Liimatainen, and Markus Pöllänen. "Attitudes and concerns on automated vehicles." Transportation research part F: traffic psychology and behaviour 59 (2018): 24-44.

[19] Hudson, John, Marta Orviska, and Jan Hunady. "People's attitudes to autonomous vehicles." Transportation research part A: policy and practice 121 (2019): 164-176.

[20] Shabanpour, Ramin, et al. "Eliciting preferences for adoption of fully automated vehicles using best-worst analysis." Transportation research part C: emerging technologies 93 (2018): 463-478.

[21] Nazari, Fatemeh, Mohamadhossein Noruzoliaee, and Abolfazl Kouros Mohammadian. "Shared versus private mobility: Modeling public interest in autonomous vehicles accounting for latent attitudes." Transportation Research Part C: Emerging Technologies 97 (2018): 456-477.

[22] Panagiotopoulos, Ilias, and George Dimitrakopoulos. "An empirical investigation on consumers' intentions towards autonomous driving." Transportation research part C: emerging technologies 95 (2018): 773-784.

[23] Rahman, Md Mahmudur, et al. "Assessing the utility of TAM, TPB, and UTAUT for advanced driver assistance systems." Accident Analysis & Prevention 108 (2017): 361-373.

[24] Buckley, Lisa, Sherrie-Anne Kaye, and Anuj K. Pradhan. "Psychosocial factors associated with intended use of automated vehicles: A simulated driving study." Accident Analysis & Prevention 115 (2018): 202-208.

[25] Hartwich, Franziska, et al. "The first impression counts–A combined driving simulator and test track study on the development of trust and acceptance of highly automated driving." Transportation research part F: traffic psychology and behaviour 65 (2019): 522-535.

[26] Kaye, Sherrie-Anne, et al. "Users' acceptance of private automated vehicles: A systematic review and meta-analysis." Journal of safety research 79 (2021): 352-367.

[27] Buckley, Lisa, Sherrie-Anne Kaye, and Anuj K. Pradhan. "A qualitative examination of drivers' responses to partially automated vehicles." Transportation research part F: traffic psychology and behaviour 56 (2018): 167-175.

[28] Körber, Moritz, Eva Baseler, and Klaus Bengler. "Introduction matters: Manipulating trust in automation and reliance in automated driving." Applied ergonomics 66 (2018): 18-31.

[29] Liu2, Peng, Zhigang Xu, and Xiangmo Zhao. "Road tests of self-driving vehicles: Affective and cognitive pathways in acceptance formation." Transportation research part A: policy and practice 124 (2019): 354-369.

[30] Yoo, Sunbin, and Shunsuke Managi. "To fully automate or not? Investigating demands and willingness to pay for autonomous vehicles based on automation levels." IATSS Research 45.4 (2021): 459-468.

[31] Deb, Shuchisnigdha, et al. "Development and validation of a questionnaire to assess pedestrian receptivity toward fully autonomous vehicles." Transportation research part C: emerging technologies 84 (2017): 178-195.

[32] Hulse, Lynn M., Hui Xie, and Edwin R. Galea. "Perceptions of autonomous vehicles: Relationships with road users, risk, gender and age." Safety science 102 (2018): 1-13.

[33] Hohenberger, Christoph, Matthias Spörrle, and Isabell M. Welpe. "How and why do men and women differ in their willingness to use automated cars? The influence of emotions across different age groups." Transportation Research Part A: Policy and Practice 94 (2016): 374-385.

[34] Davis, Fred D. "Perceived usefulness, perceived ease of use, and user acceptance of information technology." MIS quarterly (1989): 319-340.

[35] Davis, Fred D., Richard P. Bagozzi, and Paul R. Warshaw. "User acceptance of computer technology: A comparison of two theoretical models." Management science 35.8 (1989): 982-1003.

[36] Davis, Fred D., and Viswanath Venkatesh. "A critical assessment of potential measurement biases in the technology acceptance model: three experiments." International journal of human-computer studies 45.1 (1996): 19-45.

[37] Lee, Jihye, et al. "Autonomous vehicles can be shared, but a feeling of ownership is important: Examination of the influential factors for intention to use autonomous vehicles." Transportation Research Part C: Emerging Technologies 107 (2019): 411-422.

[38] Yingyu Zhang, et al. "The Research on Purchasing Intention of Fresh Agricultural Products under O2O Mode Based on the Framework of Perceived Benefits-Perceived Risk." Science in China Series F Information Sciences 6 (2015): 128-138.

[39] Xu, Zhigang, et al. "What drives people to accept automated vehicles? Findings from a field experiment." Transportation research part C: emerging technologies 95 (2018): 320-334.

[40] Helbling, Thomas. "What are externalities." Finance & development 47.4 (2010): 198.

[41] Yiqing Tang. "Consumer behavior research on positive and negative effects of the private car market in China.".2020. Southwest Jiaotong University, PhD dissertation.

[42] Bartels, Jos, and Karen Hoogendam. "The role of social identity and attitudes toward sustainability brands in buying behaviors for organic products." Journal of Brand Management 18.9 (2011): 697-708.

[43] Bauer, Raymond A. "Consumer behavior as risk taking." Proceedings of the 43rd National Conference of the American Marketing Assocation, June 15, 16, 17, Chicago, Illinois, 1960. American Marketing Association, 1960.

[44] Huang, Tianyang. "Psychological factors affecting potential users' intention to use autonomous vehicles." PLoS one 18.3 (2023): e0282915.

[45] Huang, Tianyang. "Research on the use intention of potential designers of unmanned cars based on technology acceptance model." PLoS one 16.8 (2021): e0256570.

[46] Deichmann, Johannes. Autonomous Driving's Future: Convenient and Connected. McKinsey, 2023.

[47] Krishna, G. "Understanding and identifying barriers to electric vehicle adoption through thematic analysis." Transportation Research Interdisciplinary Perspectives 10 (2021): 100364.

[48] Chamata, Johnny, and Jonathan Winterton. "A conceptual framework for the acceptance of drones." The International Technology Management Review 7.1 (2018): 34-46.

[49] Featherman, Mauricio S., and Paul A. Pavlou. "Predicting e-services adoption: a perceived risk facets perspective." International journal of human-computer studies 59.4 (2003): 451-474.

[50] Liu3, Peng, Run Yang, and Zhigang Xu. "Public acceptance of fully automated driving: Effects of social trust and risk/benefit perceptions." Risk Analysis 39.2 (2019): 326-341.

[51] Anderson, James M., et al. Autonomous vehicle technology: A guide for policymakers. Rand Corporation, 2014.

[52] Howard, Daniel, and Danielle Dai. "Public perceptions of self-driving cars: The case of Berkeley, California." Transportation research board 93rd annual meeting. Vol. 14. No. 4502. Washington, DC: The National Academies of Sciences, Engineering, and Medicine, 2014.

[53] Ullman, Jodie B., and Peter M. Bentler. "Structural equation modeling." Handbook of Psychology, Second Edition 2 (2012).

[54] Hair, Joseph F., et al. "Multivariate data analysis: Pearson new international edition." Essex: Pearson Education Limited 1.2 (2014).

[55] Fornell, Claes, and David F. Larcker. "Evaluating structural equation models with unobservable variables and measurement error." Journal of marketing research 18.1 (1981): 39-50.

[56] Anderson, James C., and David W. Gerbing. "Structural equation modeling in practice: A review and recommended two-step approach." Psychological bulletin 103.3 (1988): 411.

[57] Bentler, Peter M., and Douglas G. Bonett. "Significance tests and goodness of fit in the analysis of covariance structures." Psychological bulletin 88.3 (1980): 588.

[58] Hatcher, Larry, and Norm O'Rourke. A step-by-step approach to using SAS for factor analysis and structural equation modeling. Sas Institute, 2013.

[59] Dion, Paul A. "Interpreting structural equation modeling results: A reply to Martin and Cullen." Journal of business ethics 83 (2008): 365-368.

[60] Lee, Ju-Yeon, **-Hyuk Chung, and Bongsoo Son. "Analysis of traffic accident size for Korean highway using structural equation models." Accident Analysis & Prevention 40.6 (2008): 1955-1963.30

[61] Lee, Joo Hwan, Beom Suk **, and Yonggu Ji. "Development of a Structural Equation Model for ride comfort of the Korean high-speed railway." International Journal of Industrial Ergonomics 39.1 (2009): 7-14.

[62] Kim, Karl, Pradip Pant, and Eric Yamashita. "Measuring influence of accessibility on accident severity with structural equation modeling." Transportation research record 2236.1 (2011): 1-10.

[63] Chin, Wynne W. "Commentary: Issues and opinion on structural equation modeling." MIS quarterly (1998): vii-xvi.

[64] Gefen, David, Detmar Straub, and Marie-Claude Boudreau. "Structural equation modeling and regression: Guidelines for research practice." Communications of the association for information systems 4.1 (2000): 7.

[65] Hair, Joe, et al. "An updated and expanded assessment of PLS-SEM in information systems research." Industrial management & data systems 117.3 (2017): 442-458.

[66] Hein, Daniel, et al. "What drives the adoption of autonomous cars?." (2018).

[67] Choi, Jong Kyu, and Yong Gu Ji. "Investigating the importance of trust on adopting an autonomous vehicle." International Journal of Human-Computer Interaction 31.10 (2015): 692-702.

[68] Hohenberger, Christoph, Matthias Spörrle, and Isabell M. Welpe. "Not fearless, but self-enhanced: The effects of anxiety on the willingness to use autonomous cars depend on individual levels of self-enhancement." Technological Forecasting and Social Change 116 (2017): 40-52.

[69] Ward, Carley, et al. "Acceptance of automated driving across generations: The role of risk and benefit perception, knowledge, and trust." Human-Computer Interaction. User Interface Design, Development and Multimodality: 19th International Conference, HCI International 2017, Vancouver, BC, Canada, July 9-14, 2017, Proceedings, Part I 19. Springer International Publishing, 2017.

[70] Waytz, Adam, Joy Heafner, and Nicholas Epley. "The mind in the machine: Anthropomorphism increases trust in an autonomous vehicle." Journal of experimental social psychology 52 (2014): 113-117.

[71] Mordue, Greig, Anders Yeung, and Fan Wu. "The looming challenges of regulating high level autonomous vehicles." Transportation research part A: policy and practice 132 (2020): 174-187.

[72] Lee, Jihye, et al. "Autonomous vehicles can be shared, but a feeling of ownership is important: Examination of the influential factors for intention to use autonomous vehicles." Transportation Research Part C: Emerging Technologies 107 (2019): 411-422.

[73] Nordhoff, Sina, et al. "User acceptance of automated shuttles in Berlin-Schöneberg: A questionnaire study." Transportation Research Part F: Traffic Psychology and Behaviour 58 (2018): 843-854.

[74] Paddeu, Daniela, Graham Parkhurst, and Ian Shergold. "Passenger comfort and trust on first-time use of a shared autonomous shuttle vehicle." Transportation Research Part C: Emerging Technologies 115 (2020): 102604.

[75] Silvestri F, De Fabiis F, Coppola P. Consumers' expectations and attitudes towards owning, sharing, and riding autonomous vehicles[J]. Case Studies on Transport Policy, 2024, 15: 101112.